\documentstyle [11pt]{article}
\def\p{\partial}

\def\e{\epsilon} 
 \def\d{\delta}

\def\t{\theta} 
 \def\a{\alpha} \def\l{\lambda}
\def\f{\varphi}  \def\O{\Omega}
\def\wt{\widetilde}
\def\G{{\cal G}}\def\M{{\cal M}}
\def\C{{\cal C}}
\def\half{{1\over 2}}

\def\be{\begin{equation}}
\def\r#1{(\ref{#1})} \def\la#1{\label{#1}}                                     
\def\c#1{\cite{#1}}
\def\ee{\end{equation}}
\def\arr{\begin{array}{rll}}
\def\ea{\end{array}}
\def\bea{\begin{eqnarray}}
\def\eea{\end{eqnarray}}
\def\bealph#1
 {\setcounter{equation}{0}\renewcommand\theequation{#1\alph{equation}}\bea}
\def\eealph#1
 {\eea\setcounter{equation}{#1}\renewcommand\theequation{\arabic{equation}}}
\font\sqi=cmssq8 
\def\CC{\kern2pt {\hbox{\sqi I}}\kern-4.2pt\rm C}

\begin{document}
\rightline{hep-th/9611081}
\vskip 1 true cm                                                               
{\Large
\centerline{Hidden Symmetries of the Principal Chiral Model Unveiled} 
\vskip 1 true cm
\centerline{C. Devchand$^{1,2}$\ and\  Jeremy Schiff$^1$}
}
{\small
\vskip 0.5 true cm
\centerline{$^1$ Department of Mathematics and Computer Science}
\centerline{Bar--Ilan University, Ramat Gan 52900, Israel}
\centerline{and}
\centerline{$^2$ International Centre for Theoretical Physics}
\centerline{34100 Trieste, Italy } \centerline{ }
}
\vskip 1 true cm

\noindent {\bf Abstract.} 
By relating the two-dimensional U(N) Principal Chiral Model to a simple 
linear system we obtain a free-field 
parametrisation of solutions. Obvious symmetry transformations on the 
free-field data give symmetries of the model. In this way all  known
`hidden symmetries' and  B\"acklund transformations,
as well as a host of new symmetries, arise.

\vskip 1 true cm

\section{Introduction}

The definition of {\it complete integrability} for field theories remains
rather imprecise. One usually looks for structures analogous to those
existing in completely integrable hamiltonian systems with finitely many
degrees of freedom, such as a Lax--pair representation or conserved
quantities equal in number to the number of degrees of freedom. A very
transparent notion of integrability is that completely integrable
nonlinear systems are actually simple linear systems in disguise. For
example, the Inverse Scattering Transform for two dimensional integrable
systems such as the KdV equation establishes a correspondence between the
nonlinear flow for a potential and a constant--coefficient linear flow
for the associated scattering data. Similarly, the twistor transform for
the self-dual Yang-Mills equations converts solutions of nonlinear
equations to holomorphic data in twistor space; and for the KP hierarchy
Mulase has explicitly proven complete integrability by performing a
transformation to a constant--coefficient linear system \c{m}. In all
these examples, a map is constructed between solutions of a simple,
automatically--consistent linear system and the nonlinear system in
question. This is distinct from the Lax--pair notion of linearisation,
with the nonlinear system in question arising as the consistency condition
for a linear system. 

Just as the dynamics of completely integrable systems gets trivialised in
an auxiliary space, it seems that the confusing plethora of symmetry
transformations of these systems arise naturally from obvious
transformations on the initial data of the associated linear systems. This
idea has been exploited recently by one of us \c{j} for the KdV hierarchy:
A linearisation of KdV, mimicking Mulase's for the KP hierarchy, was used to
give a unified description of all known symmetries. 

The central feature of Mulase's construction is a group $G$ on which the
relevant linear flow acts. The group $G$ (or at least a dense subset
thereof) is assumed to be factorisable into two subgroups $G_+$ and $G_-$.
For the KP hierarchy $G$ is a group of psuedo-differential operators. For
KdV and for the two-dimensional Principal Chiral Model (PCM), as we shall
see in this paper, $G$ is a `loop group' of smooth maps from a contour
$\C$ in the complex $\l$ plane to some group H. This has subgroups $G_-$
(resp. $G_+$) of maps analytic inside (resp. outside) $\C$.  Mulase notes
that any flow on $G$ induces flows on $G_{\pm}$, but the flows on the
factors induced by a simple linear flow on $G$ can be complicated and
nonlinear. This is the genesis of nonlinear integrable hierarchies;
complete integrability is just a manifestation of the system's linear
origins. The universality of this kind of construction was noticed
by Haak {\em et al} \c{haa}.

We consider on $G$ the linear system
\be  d\ U = \O\  U,  \la{lin}\ee
where $d$ is the exterior derivative on the base space  $\M$  of 
the hierarchy, U is a $G$-valued function on  $\M$  and 
$\O$ a 1-form on $\M$ with values in $G_+$. Consistency (Frobenius 
integrability) of this system requires $ d\O = \O\wedge \O$.
In fact for KP, KdV and PCM we have  the stronger condition 
$ d \O = \O\wedge \O =0$, and \r{lin} has general solution
\be  U = e^M U_0\  ;\quad dM=\O,\quad U_0\in G. \ee
The initial data $U_0$ determines a solution of the linear system, and
hence a solution of the associated nonlinear hierarchy. A hierarchy is
specified by a choice of $G$ with a factorisation and a choice of one-form
$\O$. 

The purpose of this paper is to provide a description of the
two-dimensional Principal Chiral Model in the general framework of
Mulase's scheme. We show that for the appropriate group G, and a choice of
one-form $\O$ within a certain class, solutions of eq. \r{lin} give rise
to solutions of PCM. Thus there is a map giving, for each allowed choice
of $\O$ and each choice of initial data $U_0$, a solution of PCM. The
allowed choices of $\O$ are parametrised by free fields. The known hidden
symmetries and B\"acklund transformations of PCM all have their origins in
natural field-independent transformations of $U_0$. We also reveal other
symmetries, corresponding to other transformations of $U_0$ as well as to
transformations of the free fields in $\O$. 

We were motivated to reconsider the symmetries of PCM by a recent paper of
Schwarz \c{s}, in which infinitesimal hidden symmetries were reviewed.
However the mystery surrounding their origin remained. Further,
Schwarz's review did not encompass the work of Uhlenbeck \c{u} or previous
work on finite B\"acklund transformations \c{h}. We wish to present all
these results in a unified framework and to lift the veil obscuring the
nature of these symmetries. 

\section{The Principal Chiral Model}

The defining equations for the U(N) PCM on two-dimensional Minkowski space
$\M$ with (real) light-cone coordinates $x^+ , x^-$ are
\be\arr \p_- A_+ &=& \half [ A_+ , A_- ] ,\\[5pt]
        \p_+ A_- &=& \half [ A_- , A_+ ] ,                                  
\la{pcm}\ea\ee
where $A_\pm$ take values in the Lie algebra of U(N), i.e. they are
$N\times N$ antihermitian matrices. 
Considering the sum and difference of the two equations in \r{pcm} yields
the alternative `conserved current' form of the PCM equations 
\be \p_- A_+ + \p_+ A_-   = 0\   ,\la{pcmcc}\ee
together with the zero-curvature condition
\be \p_- A_+ - \p_+ A_- + [ A_- , A_+ ] = 0\   .\la{pcmzc}\ee
The latter has pure--gauge solution
\be  A_\pm = g^{-1} \p_\pm g\  ,\la{pg}\ee
where $g$ takes values in U(N).
Substituting this into \r{pcmcc} yields the familiar harmonic map equation
 \be  \p_-(g^{-1} \p_+ g) + \p_+(g^{-1} \p_- g) = 0.\la{pcmg}\ee
This is manifestly invariant under the  `chiral' transformation
$g\mapsto a\ g\ b$, for $a$ and $b$ constant U(N) matrices.
At some fixed 
point $x_0$ in space-time, we may choose $g(x_0) =I$, the identity 
matrix. The chiral symmetry then reduces to
\be g\mapsto b^{-1}\ g\ b .\la{un}\ee
There is a further invariance of the equations under the transformation
\be   g\mapsto g^{-1}\   .\la{inv}\ee
Eq. \r{pcm} has obvious solutions \c{zm} 
\be A_+ = A(x^+)\ ,\quad A_- = B(x^-)\ ,\la{triv}\ee 
respectively left- and right--moving diagonal matrices, i.e. taking values
in the Cartan subalgebra. (This type of solution is familiar from WZW
models and for commuting matrices the equations \r{pcm} indeed reduce to
WZW equations). In greater generality, the PCM equations imply that the
spectrum of $A_+$ (resp. $A_-$) is a function of $x^+$ (resp. $x^-$)
alone. Thus general solutions take the form: 
\be\arr  A_+ &=& s_0(x^+,x^-) A(x^+) s_0^{-1}(x^+,x^-) \\[5pt]
         A_- &=& \wt s_0(x^+,x^-) B(x^-) \wt s_0^{-1}(x^+,x^-),
\la{s0}\ea\ee 
where $A(x^+)$ and $B(x^-)$ are antihermitian diagonal matrices,
and $s_0(x^+,x^-),\wt s_0(x^+,x^-)$ are unitary. For given
$A(x^+),B(x^-)$, we have seen that there exists at least one such
solution, that with $s_0=\wt s_0=I$. We shall see in the next section that
a solution $A_\pm$ of the PCM is determined by the diagonal matrices
$A(x^+)$ and $B(x^-)$, together with another free field; and our
construction leads to solutions of precisely the form \r{s0}. Moreover, we
shall prove in section \ref{syms-proof} that hidden symmetries and
B\"acklund transformations act on the space of solutions with given
$A(x^+)$ and $B(x^-)$. 

\section{Construction of solutions}

In this section we give the formulation of the PCM in the framework of
Mulase's general scheme. Let us begin by defining a one-form on
two-dimensional Minkowski space  $\M$  with coordinates $(x^+,x^-)$, 
\be \O = -\ {A(x^+)\over 1+\l} dx^+\  -\  {  B( x^-)\over 1-\l}dx^-\  
  .\la{o}\ee
Here $A(x^+),B(x^-)$ are arbitrary diagonal antihermitian matrices, 
depending only on $x^+,x^-$ respectively. Clearly,
\be d\O\  =\ \O\wedge \O\ =\  0\  ,\ee
so that the linear equation
\be  d\ U = \O\  U  \la{l}\ee 
is manifestly Frobenius--integrable. The general solution is
\bea  U(x^+,x^-,\l) &=& e^{M(x^+,x^-,\l)}\  U_0(\l)\  ;\la{solU}\\[8pt]
  M(x^+,x^-,\l) &=& -{1\over 1+\l}\int^{x^+}_{x_0^+} A(y^+) dy^+\
               -{1\over 1-\l}\int^{x^-}_{x_0^-} B( y^-)dy^-\ 
	       ,\nonumber\eea
where $U_0$, the initial condition, is a free (unconstrained) element
of the group $G$ in  which $U$ takes values. We need to specify this group.

\noindent{\it Remarks.}

1) Since $A,B$ are anti-hermitian,  hermitian--conjugation of \r{l} yields
$$ d U(\l)^\dag = - U(\l)^\dag \O(\l^*) ,$$ whereas $U^{-1}$ satisfies
$$ d U^{-1}(\l) = - U^{-1}(\l) \O(\l) .$$ We therefore obtain the condition
\be  U^\dag(\l^*) = U^{-1}(\l) .\la{hc}\ee

2) $\O$ has poles at $\l=\pm1$, so it is analytic everywhere in the
$\l$-plane including the point at $\infty$,
except in two discs with centres at $\l=\pm1$. We therefore
introduce  a contour $\C$, the union of
two small contours $\C_\pm$ around $\l=\pm1$ (such
that $\l=0$ remains outside both of them), dividing the $\l$-plane into
two distinct regions: the `outside' $\{|\l-1|>\d\}\cap \{|\l+1|>\d\}$ 
and the `inside' $\{|\l-1|< \d\}\cup \{|\l+1|<\d\}$, where $\d<1$ is
some small radius. 

\noindent{\it Definition.} $G$ is the group of smooth maps $V=V(\l)$ 
{}from the contour $\C$ to $GL(N,\CC)$ satisfying the condition 
$V^\dag(\l^*) = V^{-1}(\l)$.

We are going to pretend that there exists a Birkhoff factorisation 
$G = G_-\ G_+$, where $G_-$ denotes the group of maps analytic inside 
$\C$ and $G_+$ denotes the group of maps analytic outside $\C$ {\it and} 
equal to the identity at $\l=\infty$. The corresponding Lie algebra
decomposition is~ $\G = \G_-\ \oplus\ \G_+$. This factorisation is
definitely a pretence; but the point is that sufficiently many elements
of $G$ do factor this way so that the results we will obtain using this
factorisation do hold. For a more precise discussion we refer to \c{u,haa}.

We now have the spaces in which the objects in \r{l},\r{solU} take values.
Clearly, $\O$ is a one-form on $\M$ with values in the Cartan subalgebra
of the Lie algebra $\G_+$. The matrix $U = U(x^+,x^-,\l)$ is a map from
$\M$ to $G$ and $U_0(\l)$ is an element of $G$ (independent of $x^\pm$). 

Consider a solution $U$ of \r{l}. Assuming the existence of a
Birkhoff factorisation for  $U$, we can write  
\be  U = S^{-1}\ Y\   .\la{split}\ee
where $S^{-1}: \M \rightarrow G_-$  and $Y:\M \rightarrow G_+$.
Now, applying the exterior derivative on both sides and using \r{l} yields
\be S\O S^{-1} = - d S S^{-1} + d Y Y^{-1}  .\ee
$S\O S^{-1}$, which takes values in the Lie algebra $\G$,
decomposes into its components in the  $\G_-$ and $\G_+$ 
subalgebras.  The above equation allows us 
to write separate equations for the projections:
\be\arr  (S\O S^{-1} )_- &=& - d S S^{-1}   \\[5pt]
         (S\O S^{-1} )_+ &=&   d Y Y^{-1}  .\la{Y}\ea\ee
Here the suffix notation denotes the projection of an element 
of $\G$ into $\G_\pm$. We introduce a one-form $Z$ taking values in $\G_+$,
\be   Z =   d Y Y^{-1} =  (S\O S^{-1} )_+   \la{Zdef}.\ee
Now, since $S$ takes values in $G_-$, it is analytic at
$\l=\pm 1$ and has two power-series representations, converging in discs
with centres at $\l=\pm 1$, viz.
\be  S = \sum_{n=0}^\infty s_n(x^+,x^-) (1 + \l )^n 
       = \sum_{n=0}^\infty \wt s_n(x^+,x^-) (1 - \l )^n   ,\la{expS}\ee
where the coefficients $s_0(x^+,x^-), \wt s_0(x^+,x^-)$ are U(N)-valued 
matrices. Inserting these expansions in $(S\O S^{-1})$, we see that 
only the $s_0$ and $\wt s_0$ terms survive the projection to the $\G_+$ 
subalgebra, yielding 
\be  Z= (S\O S^{-1} )_+\ =\  - {s_0 A(x^+) s_0^{-1}\over 1+\l}dx^+
                  - {\wt s_0 B(x^-) \wt s_0^{-1}\over 1-\l}dx^-. \la{Z}\ee
Define
\be A_+ = s_0 A(x^+) s_0^{-1}\  ,\quad 
    A_- = \wt s_0 B(x^-) \wt s_0^{-1} .\la{A}\ee

\vskip 0.3cm \noindent
{\it These satisfy the PCM equations \r{pcm}.}

\vskip 0.3cm \noindent
The proof is immediate. From \r{Zdef}
\be  d\ Z =\  Z \wedge Z  .\la{ZZ}\ee
Inserting the form \r{Z} in this equation yields 
$$ { \p_+ A_- \over 1-\l} - {\p_- A_+\over 1+\l} 
 + \half \left({1\over 1-\l} -{1\over 1+\l}\right)[ A_+ , A_- ] =\ 0\ .$$
Since Y takes values in $G_+$, for consistency this equation needs to hold
for all values of $\l$ away from $\pm 1$. In other words, the coefficients
of ${1\over 1-\l}$ and ${ 1\over 1+\l }$ must be separately zero. This 
yields precisely the two equations in \r{pcm} as integrability conditions.

Note that the solutions \r{A} have precisely the form \r{s0}.
We have seen that for given diagonal matrices $A(x^+)$ and  $B(x^-)$,
{\it a solution of the linear field--independent system \r{l} 
determines a solution of the PCM in the spectral class of $A$ and $B$.}

In fact the general solution of \r{l} takes the form \r{solU}, where the
$e^M$ factor contains only spectral information (i.e. $A,B$). Everything
else is encoded in the free element $U_0(\l)\in G$. So the
freely--specifiable data $\{ A(x^+),B(x^-),U_0(\l)\}$ corresponds to a
solution of the PCM. Given any choice of these three fields, a solution
of the PCM can be constructed in the following stages:

(a) Construct the corresponding $U(x^+,x^-,\l)$ from \r{solU}.

(b) Perform the factorisation \r{split} to obtain $S(x^+,x^-,\l)$.

(c) Perform the two expansions \r{expS} to extract the coefficients
$s_0(x^+,x^-)$ and $\wt s_0(x^+,x^-)$.

(d) Insert these in \r{A} to obtain a solution of the PCM.

Note that this procedure is purely algebraic, though the factorisation may
not be very easy to perform in practice. However, it is clear that for
any choice of $A(x^+),B(x^-)$ (which is tantamount to fixing the
spectral class of $A_\pm$), every $U_0(\l)\in G$ corresponds to a solution
of the PCM. In fact there is a large redundancy, for a
right--multiplication
 \be U_0\mapsto U_0 k_+\ ;\quad k_+ \in G_+ \la{rk+}\ee
corresponds to a right-multiplication $U\mapsto U k_+$, which does nothing
to alter the $S^{-1}$ factor in \r{split}. PCM solutions therefore
correspond to $G_+$ orbits in $G$, or equivalently, $U_0(\l)$'s from the
Grassmannian $G/G_+$. This correspondence is, however, still redundant: 
Consider a left--multiplication by a diagonal matrix analytic inside $\C$,
 \be U_0\mapsto h_-\ U_0\ ;\quad h_- \in G_{0,-}\  ,
     \mbox{ the maximal torus of } G_- .\la{lh-}\ee
Since this commutes with the diagonal $e^M$, it corresponds to a
transformation $S^{-1}\mapsto h_-\ S^{-1} $. However, since $h_-$ is a
diagonal matrix, the $A_\pm$ in \r{A} do not notice this transformation;
they are invariant. The correct space of $U_0$'s corresponding to
solutions of \r{pcm} in each spectral class of $A_\pm$ is therefore the
double coset $G_{0,-}\backslash G/G_+ $. In particular, natural
transformations of $U_0(\l)$ preserving this double coset correspondence
induce symmetry transformations on the space of PCM solutions.

\section{The extended solution}\la{exts}

The fact that the consistency condition \r{ZZ} with $Z$ given by \r{Z} 
yields the PCM equations is well known. Writing \r{Zdef}
in more familiar form,
$$ d\ Y =\  Z\ Y\    ,$$
it is precisely the PCM Lax-pair \c{p,zm},
\be\arr
\left( \p_+ + {1\over 1+\l}A_+\right) Y &=& 0\\[5pt]
\left( \p_- + {1\over 1-\l}A_-\right) Y &=& 0\  .\la{lax}\ea\ee
It is easy to check that the $Y$ we have defined above has all the 
properties required of a solution of this pair of equations:

1.  As a function of $\l$, the only singularities of $Y$ on the entire 
$\l$-plane including the point at $\infty$ are at $\l=\pm 1$.

2. The solution of the system \r{lax} is easily seen to satisfy the reality
condition \r{hc}
\be  Y^\dag(\l^*) = Y^{-1}(\l) .\ee

3. There is an invariance of the Lax system:
$Y(x,\l) \mapsto Y(x,\l) f(\l)$, which is usually fixed by setting
\be Y(x_0,\l) =I\  ,\la{Yx0}\ee for some fixed point $x_0$.  
This invariance corresponds to  right--multiplications \r{rk+} of $U_0$ 
and the condition \r{Yx0} corresponds to choosing a representative point
on the $G_+$ orbit of $U_0$ in $G$. 

4. At $\l=\infty$, $\p_+ Y = \p_- Y = 0 $, so $Y(x, \l=\infty)$ is a constant
and using \r{Yx0} we obtain
\be Y(x, \l=\infty) = I .\la{Yinf}\ee

5. The system \r{lax} yields the expressions
\be  A_+ = (1+\l) Y \p_+ Y^{-1}\  ,\quad A_- = (1-\l) Y \p_- Y^{-1} 
  ,\la{Aext}\ee
which together with \r{Yx0} and \r{pg} imply that
\be  Y(x, \l=0) = g^{-1}  .\la{Y0}\ee

We have already seen that the $A_\pm$ solving \r{pcm} may be recovered
{}from power series expansions around $\l=\pm 1$ of the $S^{-1}$ factor of
$U$ using the expressions \r{A}. We now see that solutions may equally be
obtained from the $Y$ factor using \r{Y0} and \r{pg}. We can also obtain
solutions from the $Y$ factor by expanding around $\l =\infty$. Denoting
the leading terms consistently with \r{Yinf}, 
\be Y(x,\l) = I + {f(x)\over \l}+\dots ,\la{linf}\ee where $f(x)$ is 
antihermitian, the $\l =\infty$ limit of \r{Aext} yields the expressions
\be A_\pm = \mp \p_\pm f\  ,\la{Af}\ee
which identically satisfy \r{pcmcc} and shift the dynamical description to
\r{pcmzc} instead, which acquires the form
\be \p_- \p_+ f + \half [ \p_- f , \p_+ f ] = 0\  .\la{pcmf}\ee
This equation is known as the `dual formulation' of the harmonic map
equation \r{pcmg}. A $Y(x,\l)$ obtained from the factorisation procedure
automatically yields a solution of this equation on expansion around
$\l=\infty$. We therefore see that the factorisation \r{split} produces a
$Y(x,\l)$ which interpolates between the dual descriptions of PCM
solutions; yielding a U(N)--valued solution $g^{-1}$ of the equation
\r{pcmg} on evaluation at $\l=0$ and a Lie-algebra-valued solution $f$ of
the alternative equation \r{pcmf} on development around $\l=\infty$. The
$G_+$--valued $Y(x,\l)$ thus encapsulates these dual descriptions of
chiral fields and this field was aptly named the {\it extended solution}
of the PCM by Uhlenbeck \c{u}.

We shall later need information about the next-to-leading-order term
in the expansion of $Y$ around $\l=0$. If we substitute
\be Y = (I + \l\f) g^{-1} + O(\l^2) ,\la{p} \ee
where $\f$ is a Lie-algebra-valued field, into \r{Aext}, and use \r{pg},
we obtain the following first-order equation for $\f$:
\be  \p_\pm \f + [A_\pm \ ,\ \f]  = \pm A_\pm.\la{peq}\ee 
The consistency condition for this is just \r{pcmcc}.
 
Reflecting the $G_+$--valued extended solution $Y(x,\l)$, there is also
the $G_-$--valued $S(x,\l)$, which clearly also describes some extension
of the PCM solution given by the expression \r{A}. Using 
$dS S^{-1} = -(S\O S^{-1})_- = -(S\O S^{-1}) + (S\O S^{-1})_+$ 
we find the following flows for the components of $S$, which we shall 
need later:
\begin{eqnarray}
\p_+ s_n &=&s_{n+1} A - A_+ s_{n+1} \la{s1}\\
\p_- s_n &=&\sum_{r=0}^n \frac{s_r B - A_- s_r }{2^{n-r+1}}\\
\p_+ \wt s_n &=&\sum_{r=0}^n \frac{\wt s_r A - A_+ \wt s_r}{2^{n-r+1}}\\
\p_- \wt s_n &=&\wt s_{n+1} B - A_- \wt s_{n+1}\la{s4}.
\end{eqnarray}
Using \r{A} and these equations for $n=0$ yields the interesting flow equations
\be\arr \p_+ A_+  &=&  s_0\ \p_+ A\  s_0^{-1}\  +\ 
                      [A_+ , [A_+\ ,\ s_1 s_0^{-1} ] ] \\[5pt]
  \p_- A_- &=&  \wt s_0\ \p_- B\  \wt s_0^{-1}\  +\ 
                      [A_- , [A_-\ ,\ \wt s_1 \wt s_0^{-1} ] ].  \la{b}\ea\ee

\section{Symmetry transformations unveiled}\la{syms}

Non-space-time symmetry transformations of the PCM were traditionally
derived using main\-ly 
guesswork inspired by analogies with other integrable
models like the sine-Gordon model. Their origin remained largely veiled in
mystery and they were therefore called `hidden symmetries'. Previous
discussions of them have recently been reviewed by Schwarz \c{s} and
Uhlenbeck \c{u}. In the framework of the
present paper there is nothing `hidden' about these symmetries.
As we shall see, in terms of the
free-field data $U_0(\l), A(x^+),B(x^-)$, the veil hiding these
symmetries is entirely lifted: the most natural field-independent
transformations of these free fields, which preserve their analyticity
properties in their respective independent variables, induce the entire
array of known symmetry transformations of PCM fields and more. Moreover,
the algebraic structure of the symmetry transformations is completely
transparent when acting on the free-field data, and there is no need 
to compute commutators and check closure using the complicated
action of the symmetries on physical fields. The physical fields 
automatically carry representations of all the symmetry actions on the 
free-field data.

In this section we classify PCM symmetry transformations according to the
corresponding transformations of the free fields. The formulas for the 
induced transformations on the extended solutions $Y$, on the chiral fields 
$g$ and on the potentials $A_\pm$ will be derived in the next section.

\subsection{Symmetry transformations of $U_0$}

We first list symmetry transformations which leave $A(x^+)$ and $B(x^-)$
unchanged.

\subsubsection{Right dressings}

Right-actions by elements of the $G_+$ subgroup \r{rk+} have already been
seen to correspond to trivial redundancies and have already been factored 
out. This leaves the possibility of right--multiplying $U_0$ by an
element of $G_-$, 
\be U_0\mapsto U_0 k_-\ ;\quad k_- \in G_-\  .\la{rk-}\ee
Such transformations fall into the following classes:

{\bf a)} $k_- = b$, a constant (i.e. an element of U(N)).
This may easily be seen to induce the transformations
$Y \mapsto b^{-1} Yb$ and  $g \mapsto b^{-1} gb$,
i.e. the symmetry \r{un}.

{\bf b)} If we take 
$k_- = \left( I+ {N(\mu) \over \l - \mu} \pi \right),$
having a simple pole at a single point $\l=\mu$ outside $\C$
(here $N(\mu)$ is a $\l$-independent matrix),
the transformations induced on the chiral fields are precisely the 
B\"acklund transformations of \c{h,o}.

{\bf c)} We are presently considering the U(N) PCM. For the GL(N,$\CC$)
PCM  we could consider finite transformations with $k_-$ in a triangular 
subgroup of $G_-$. Such transformations induce the explicit transformations 
discussed by Leznov \c{l}. We will not go into details of this.

{\bf d)} General $k_-(\l)$ infinitesimally close to the identity.
This is a realisation of the algebra $\G_-$ on the free-field $U_0(\l)$
and is a remarkably transparent way of expressing the action of the 
celebrated loop algebra of hidden symmetries \c{d} of the PCM. The precise 
structure of this algebra has not been properly identified before.
 
{\bf e)} General finite $k_-(\l)$.
This finite version of the infinitesimal symmetries in d) reproduces 
(modulo some details) the loop group action on chiral fields $g$ and on 
extended maps $Y$ given by Uhlenbeck in sect.5 of \c{u}. 

\subsubsection{Left dressings}

Left actions on $U_0$ by elements of $G_{0,-}$ have already been pointed 
out to leave the associated solution of the PCM invariant (see \r{lh-}).
We wish to consider only left actions on $U_0$ that descend to the double
coset $G_{0,-}\backslash G/G_+$, i.e. actions by elements
that commute with $G_{0,-}$. Thus we  have only the transformations
\be U_0 \mapsto  h_+ U_0 ;\quad h_+ \in G_{0,+}. \la{lh+}\ee
This is the action of an infinite-dimensional abelian group,
which has  not yet appeared in the literature. The infinitesimal 
version of this gives an infinite set of mutually commuting
flows also commuting with the PCM flow. This is the PCM hierarchy.

\subsubsection{Reparametrisations of $U_0(\l)$}

These are transformations generated by $\l$-diffeomorphisms
\be U_0(\l) \mapsto U_0(\l + \e(\l))  .\ee 
General reparametrisations can move $\C_\pm$ to curves that do not
enclose $\pm 1$. The easiest way to prevent this is to restrict
the diffeomorphisms to those that fix $\pm 1$. 
For infinitesimal diffeomorphisms
this condition is not strictly necessary. It turns out however that the
infinitesimal diffeomorphisms fixing $\pm 1$  are technically 
simpler (in terms of their action on $g,Y$) and these 
give (modulo a detail that will be explained) the `half Virasoro'
algebra described in \c{s}. We show how this can be extended to a full
centreless Virasoro algebra.

The only finite reparametrisations of the $\l$-plane preserving
$\pm 1$ are 
\be U_0(\l) \mapsto U_0\left( {a \l + b\over  b\l+a} \right)\  
  ,\quad a^2 + b^2 = 1.  \ee
These induce the  $S^1$ action of sect. 7 of \c{u}.

\subsection{Symmetry transformations of $A(x^+), B(x^-)$}

We now consider symmetries that keep $U_0$ fixed. For symmetries
acting just on $A(x^+)$ it is natural to consider

\noindent a) Shifts $A(x^+)\mapsto A(x^+)+\a(x^+)$, where $\a(x^+)$ is
a diagonal antihermitian matrix.

\noindent b) Rescalings $A(x^+)\mapsto \rho(x^+) A(x^+)$ where $\rho(x^+)$ is
a scalar function.

\noindent c) Reparametrisations $A(x^+) \mapsto A(x^+ + \e(x^+))$.

\noindent
There are other possibilities. Similar symmetries exist for $B(x^-)$.
All these symmetries are new.

\subsection{Other symmetry transformations}

Two other symmetries of PCM should be mentioned. The first is a particularly
significant combination of an action on $U_0$ with an action on $A,B$.
The second is not strictly within the class of symmetries we have
been considering, as it acts on the coordinates as well as the fields. 

\subsubsection{Inversion}

The transformation
\be U_0(\l) \mapsto U_0(\l^{-1})  \quad \mbox{and} \quad
 (A,B) \mapsto (-A,-B) \ee
may easily be seen to induce the inversion symmetry \r{inv}.

\subsubsection{Lorentz transformations}

The transformation
\be\arr U_0~\mbox{invariant}, \quad A &\mapsto& \t_+A, \quad B \mapsto \t_-B 
   \\[5pt]
x^\pm &\mapsto&\t_\pm^{-1}x^\pm  \ea
\ee
induces the residual Lorentz transformations in light cone coordinates
\be 
A_\pm \mapsto \t_\pm A_\pm\  ,\quad
                          x^\pm \mapsto \t_\pm^{-1} x^\pm. \ee
We can also consider more general reparametrisations of $x^{\pm}$.

\section{Induced symmetries of PCM fields}\la{syms-proof}

As we have already claimed, natural transformations on the free--field
data, $U_0(\l),\linebreak[0]A(x^+),\linebreak[0]B(x^-)$ induce, 
through Birkhoff
factorisation, rather complicated transformations on the PCM fields
$Y(x,\l),g(x),A_\pm(x)$; and (field--independent) representations of
symmetry algebras induce (field-dependent) representations on the PCM
fields. In this section we prove this for the intereresting
and not immediately obvious
cases listed in the previous section. We also comment on
the relation with previous results in the literature. 

\subsection{Right dressings}

Consider the transformation induced by \r{rk-} on $U(x,\l)$.
\be U =\ S^{-1}\  Y\  \mapsto\ U_{new} =\  S^{-1}\ Y k_- .\la{Udd}\ee
Birkhoff factorisation of $Y k_-$ yields (in the obvious notation) 
\be U_{new} =\  S^{-1}\ (Y k_- )_- (Y k_- )_+  
      = S_{new}^{-1}\  Y_{new}  \la{Ud}\ee
In other words, we have the symmetry transformation
\be Y\  \mapsto\ (Y k_- )_+\  ,\la{Yd}\ee
which is just the representation of $G_-$ described by Uhlenbeck
in sect. 6 of \c{u} (except that she uses a subgroup of $G_-$).
We can equivalently write 
\be Y\  \mapsto\ (Y k_- Y^{-1} )_+  Y\  .\la{Ydd}\ee
Now writing $k_- = I + \e(\l)$ with  $\e(\l) \in \G_-$ an infinitesimal 
parameter, we obtain the infinitesimal version of this,
\be Y\  \mapsto  \left( I + (Y \e(\l) Y^{-1} )_+ \right) Y\  .\ee
We note that this directly gives the generating function of \c{Chand} 
for these transformations, which was originally obtained by extrapolation 
from the leading terms in a power series expansion \c{d}.
The $\G_+$ projection corresponds to taking the singular part at $\l=\pm 1$.
This may be done using a contour integral, so that
this transformation takes the form
\be Y(x,\l)\  \mapsto  \left( I +
{1 \over 2\pi i} \int_\C {Y(x,\l')\e(\l') Y^{-1}(x,\l') \over \l' - \l} d\l' 
                    \right)\  Y(x,\l)\   .\la{Ydinf}\ee
Here $\C_\pm$ are oriented counter-clockwise around $\pm 1$.
The transformation for $g$ may be read off by taking the 
$\l \rightarrow 0$ limit,
yielding the form of the transformation given in \c{un,s}, 
\be g \mapsto g\left( I - \frac1{2\pi i}\int_{\C}
\frac{Y(x,\l')\e(\l')Y^{-1}(x,\l')}{\l'} d\l'\right)\  .\la{grd}\ee

The parameter of this infinitesimal transformation, 
$\e({\l})$ is an arbitrary infinitesimal $\G_-$ element. In 
particular, if we introduce a basis $\{T^a\}$ for the Lie algebra of 
antihermitian matrices, we can take $\e({\l})$ proportional to $\l^r T^a$,
$r\in{\bf Z}$ . This gives an infinite set of transformations, which we 
denote $J_r^a$, and which satisfy the commutation relations
\be [J_r^a,J_s^b] = \sum_c f^{ab}_c J_{r+s}^c,  \la{ukm}\ee
where the $f^{ab}_c$ are the structure constants defined by 
$[T^a,T^b] = \sum_c f^{ab}_c T^c$. Although the commutation relations
of a centreless Kac-Moody algebra thus appear, this is 
{\em not} sufficient to identify the symmetry algebra $\G_-$ with a 
centreless Kac-Moody algebra. We illustrate this in two ways: 
first we show that in $\G_-$ there exist certain linear relations 
absent in a Kac-Moody algebra, and second we show that in $\G_-$
the $J_r^a$ are not a spanning set. 

The crucial point is that although we can certainly try to expand elements 
of $\G_-$ in Laurent series, and finite sums of matrices of
the form $\l^r T^a$ are certainly in $\G_-$, the natural way to expand
an element of $\G_-$ is in a Taylor series in $\l+1$ (or alternatively in
$\l-1$). Taking $\e(\l)$ in \r{grd} proportional to $(\l+1)^n T^a$, for 
$n\ge 0$, we can define a set of transformations $K_n^a$ satisfying the
relations
\be [K_n^a,K_m^b] = \sum_c f^{ab}_c K_{n+m}^c \qquad  n,m\ge 0 .\la{kkm}\ee
Considering the expansion of $\l^r$ in powers of $\l+1$ (valid in
$\vert\l+1\vert<\delta$), we find that the $J_n^a$ are expressed as 
linear combinations of the $K_n^a$ in the following way:
\be J_r^a = \left\{
                 \matrix{  \sum_{n=0}^r (-1)^{n+r}
                    \pmatrix{r\cr n\cr} K_n^a &
                     \quad\quad r\ge 0 \cr
                 \sum_{n=0}^{\infty} (-1)^r
                    \pmatrix{n-r-1\cr -r-1\cr}K_n^a &
                          \quad\quad r< 0 \cr} \right. \la{JfK} \ee
It is straightforward, using standard formulae for sums of binomial
coefficients (see for example \c{f}), to check that these linear
combinations, in virtue of \r{kkm}, imply the commutation relations \r{ukm}. 
The relation between the $J_r^a$ for non-negative $r$ can be
inverted: we find
\be K_n^a=\sum_{r=0}^n \pmatrix{n\cr r\cr}J_r^a\  .\la{KfJ}\ee
Now, if our symmetry algebra were indeed a Kac-Moody algebra with generators 
$J_r^a$ satisfying \r{ukm}, we would be able to define the algebra elements 
$K_n^a$ (which certainly exist as symmetry generators)
{}from the $J_r^a$'s with non-negative $r$ using \r{KfJ}. 
When we substitute \r{KfJ} into the infinite sum in \r{JfK} we find that we 
cannot reorder the summations to express this infinite sum as a 
linear combination of the  $J_r^a$'s with $r\ge 0$. In other words, this 
infinite sum is not in the Kac-Moody algebra. We thus have our first distinction 
between a Kac-Moody algebra and $\G_-$: In a Kac-Moody algebra the
elements $K_n^a$ and the elements $J_r^a$ for $r<0$ need to be linearly 
independent, whereas in the PCM symmetry algebra $\G_-$  they  are
linearly dependent via the relationship given in \r{JfK}.

The second distinction is that in $\G_-$, unlike in a regular Kac-Moody 
algebra, the elements $\{ J_r^a \}$  are not a spanning set. Elements of $\G_-$
need to be analytic inside $\C$. There are therefore elements of $\G_-$ that
do not have Laurent expansions in powers of $\l$; consider for example
an $\e(\l)$ proportional to $\ln\l$, defined with a cut from
$0$ to $\infty$ along half of the imaginary axis. Now, the reader may be
concerned that  we have claimed
that $\G_-$ is spanned by the $K_n^a$, that the relationship between
the $K_n^a$ and the $J_r^a$ for $r\ge 0$ is invertible, but that the $J_r^a$
(and therefore certainly the $J_r^a$ for $r\ge 0$) are not a spanning set for 
$\G_-$. There is absolutely no contradiction here. As we have seen above,
the relationship between the $K_n^a$ and the $J^r_a$ for $r\ge 0$ implies 
that finite linear combinations of the $K_n^a$ can be written as linear
combinations of the $J_r^a$ for $r\ge 0$, but for infinite linear 
combinations of the $K_n^a$ this is not the case. However, it does suggest that
we should be able in some sense to approximate elements of $\G_-$
given by infinite sums of the $K_n^a$'s by finite sums of the $J_r^a$,
which are equivalent to finite sums of the $K_n^a$. This is indeed the case,
as follows from a classical theorem
in complex analysis, Runge's theorem (see, for example, \c{r}). Runge's
theorem implies the remarkable fact that a function analytic 
on an arbitrary finite union of non-intersecting open discs can be 
approximated uniformly and to any accuracy on any closed subset of the 
union by a polynomial. In particular, this implies that 
elements in $\G_-$ can be approximated uniformly and to any
accuracy on $\{\vert \l-1\vert<\delta\}\cup\{\vert\l+1\vert<\delta\}$ by
a finite linear combination of the $J_r^a$ for $r\ge 0$. 

To conclude this section we note that the contour integral in \r{grd}
is easily evaluated when $\e(\l)$ is proportional to $\l^r$: For $r<0$
the integral is evaluated by shrinking $\C$ to a contour around $0$; for
$r>0$ to a contour around  $\infty$; and for $r=0$ to a pair of
contours around $0$ and $\infty$.

\subsection{The B\"acklund transformation}

The element $k_- \in G_-$ in \r{rk-} can clearly have all variety of
singularities {\it outside} $\C$.  Trying to give $k_-$ just one simple pole 
at the point $\l=\mu$ outside $\C$, suggests the natural form \c{zm}
\be  k_-(\l,\mu) = \left( I+ { N(\mu) \over \l - \mu}\right) .\ee
For the satisfaction of the reality condition \r{hc} for elements of
$G_-$ we require that $N^\dag = { N N^\dag \over \mu-\bar\mu}= -N $. These
conditions are satisfied by $N=(\mu-\bar\mu)\pi$, if $\pi$ is a projector
satisfying $\pi^2=\pi=\pi^\dag$. Such transformations thus correspond to
finite right-dressing transformation of the particular form
\be U_0\mapsto U_0\ \left( I+ {\mu - \bar\mu \over \l - \mu} \pi \right)
   .\la{back}\ee
Note that $k_-$ in fact has a singularity at $\l=\bar\mu$ as well, since
$(I-\pi)$ has zero determinant. 
Using \r{Udd} we obtain the transformation
\be  U\mapsto  S^{-1}\ 
\left( I+ {\mu - \bar\mu \over \l - \mu} Y(\l) \pi  Y^{-1}(\l) \right)\  
               Y(\l)\    .\la{Ub}\ee
In order to factorise the middle factor, we introduce a hermitian 
projector $P=P^\dag=P^2 $, independent of $\l$ (but not of
$x^\pm$). Using this we see that
$$\arr &&
\left(I+ {\mu - \bar\mu \over \l - \mu}\ Y(\l) \pi Y^{-1}(\l)\right)Y(\l) 
\\[5pt] &=& \left( I+ {\mu - \bar\mu \over \l - \mu}\  P \right)
\left( I+ {\bar\mu - \mu \over \l - \bar\mu}\ P \right)
\left(I+ {\mu - \bar\mu \over \l - \mu}\  Y(\l) \pi  Y^{-1}(\l) \right)Y(\l)
\\[5pt] &=& \left( I+ {\mu - \bar\mu \over \l - \mu}\ P \right)
\left( I+ {\mu - \bar\mu \over \l - \mu} \left(I-P\right) Y(\l) \pi 
+ {\bar\mu - \mu \over \l - \bar\mu}\ PY(\l)\left(I-\pi\right) \right) \ea$$
To have an acceptable factorisation, all we need now is that the right-hand
factor above be regular outside $\C$. Specifically, we require regularity
at $\mu$ and $\bar\mu$, which yields algebraic conditions relating the
projectors $P$ and $\pi$, viz.
$$ (I-P) Y_\mu \pi = 0\ ,\quad PY_\mu(I-\pi)=0  ,$$
where $Y_\mu$ denotes $Y(\l)$ evaluated at $\l=\mu$.
If we write $ \pi =  v (v^\dag v )^{-1} v^\dag $ (see \c{h}), 
these equations are solved by the expression
$$ P = 
Y_\mu v \left(v^\dag Y^\dag_\mu Y_\mu v \right)^{-1}v^\dag Y^\dag_\mu\ .$$
Now we can read-off the induced transformation rules for $Y$ and $g$.
These are just the known PCM B\"acklund transformations \c{h,o,zm}.

\subsection{Left dressings}\la{Ld}

Here we consider in detail the left dressings \r{lh+}. Matrices $h_+\in
G_{0,+}$ commute with $M$, so such transformations act by left
multiplication on $U$, i.e. $U\mapsto h_+\ U=h_+\ S^{-1}Y
=S^{-1}(Sh_+\ S^{-1})Y$. 
Hence the action on
$Y$ is given by
\be Y \mapsto (h_+ S^{-1})_+ Y = (S h_+ S^{-1})_+ Y. \ee
For an infinitesimal transformation $h_+=I+\e$, $\e\in \G_{0,+}$ and
we have
\begin{eqnarray}
 Y &\mapsto& \left(I+(S\e S^{-1})_+\right)Y \nonumber\\
   &=& \left( I +\frac1{2\pi i}\int_{\C} \frac{S(\l')\e(\l')S^{-1}(\l')}{
              \l'-\l} d\l'\right)Y,
\end{eqnarray}
implying 
\be g \mapsto g\left( I - \frac1{2\pi i}\int_{\C} 
           \frac{S(\l')\e(\l')S^{-1}(\l')}{\l'} d\l' \right). \la{ldog}\ee
In general $\e$ has the form
\be \e(\l) = \sum_{n=1}^\infty \left( 
   \frac{\alpha_n}{(1+\l)^n} + \frac{\wt \alpha_n}{(1-\l)^n} \right), 
     \ee
where the $\alpha_n,\wt \alpha_n$ are 
constant infinitesimal diagonal matrices. 
The integral in \r{ldog} is evaluated by computing the residues of the
integrand at $\l'=\pm 1$.
For example, the case $\alpha_1\not=0$ with all
other $\alpha_n,\wt\alpha_n$ zero yields the transformation rules
\begin{eqnarray}
g^{-1}\delta g &=& -s_0\alpha_1 s_0^{-1} \nonumber\\
\delta A_+ &=& \left[A_+,[s_1s_0^{-1},s_0\alpha_1 s_0^{-1}]\right]
    \la{ldoA}\\
\delta A_- &=& -{\textstyle \frac12}[A_-,s_0\alpha_1 s_0^{-1}].\nonumber
\end{eqnarray}
Similarly, if $\alpha_2\not=0$ with all other $\alpha_n,\wt\alpha_n$ 
zero we find
\begin{eqnarray}
g^{-1}\delta g &=& -\left(s_0\alpha_2s_0^{-1}+[s_1s_0^{-1},
  s_0\alpha_2s_0^{-1}]  \right)  \nonumber\\
\delta A_+ &=& \left[A_+, [s_2s_0^{-1},s_0\alpha_2s_0^{-1}]-
                          [s_1s_0^{-1},s_0\alpha_2s_0^{-1}]s_1s_0^{-1}
\right]\la{awful}\\
\delta A_- &=& -\left[A_-, {\textstyle \frac14}s_0\alpha_2s_0^{-1}
 + {\textstyle \frac12}[s_1s_0^{-1},s_0\alpha_2s_0^{-1}]\right].\nonumber
\end{eqnarray}
The formulae for $\delta A_\pm$ are computed using
the variation  of the relation \r{pg},
\be \delta A_\pm = \p_{\pm}(g^{-1}\delta g) + [A_\pm,g^{-1}\delta g]. \ee
and equations \r{s1}-\r{s4}. The latter also 
allow one to check directly that the above transformations  are indeed
infinitesimal symmetries, i.e. that $\p_-\delta A_+ + \p_+\delta A_-=0$. 

Now considering the sector of PCM in which $A=\alpha_1$, independent 
of $x^+$, we see that the $\p_+$-derivations of $A_\pm$ given by
\r{pcm} and \r{b} are effected by the transformations \r{ldoA}.
So left dressing transformations with only $\alpha_1$ non-zero
correspond to $x^+$ translations in this sector. 
Similarly the transformations \r{awful} can be seen to be related to
coordinate translations in an extended system (described in the appendix)
belonging to a {\em hierarchy} associated to the PCM. 
Whenever an infinite dimensional abelian symmetry algebra 
(like $\G_{0,+}$) is identified
in a system, it is possible to define a corresponding hierarchy. 
Traditionally, for each
generator in the algebra a coordinate is introduced and the flow in each
coordinate is defined as the infinitesimal action of the corresponding 
symmetry. In our formulation  there is an alternative 
way to define  a PCM hierarchy. Instead of working on a 
space $\M$ with coordinates $(x^+,x^-)$, we work on a larger space $\M$
with $2P$  coordinates $(x_1^+,\ldots,x_P^+,x_1^-,\ldots,x_P^-)$ and
replace the $\O$ of \r{o} by 
\be \O=-\sum_{n=1}^P \left( \frac{A_n(x_n^+)dx_n^+}{(1+\l)^n} +
                            \frac{B_n(x_n^-)dx_n^-}{(1-\l)^n} \right), 
          \la{nO}\ee
where the $A_n(x_n^+),B_n(x_n^-)$ are all antihermitian diagonal matrices,
each depending on only one coordinate. The associated nonlinear equations
are again the  equations $dZ=Z\wedge Z$, where $Z=(S\O S^{-1})_+$ and $S$
is a map from $\M$ to $G_-$. For the case $P=2$ we write out this system of 
equations in full in the appendix. Another possibility of 
obtaining a hierarchy within our framework is 
to enlarge $\M$ to a space with $2NP$ coordinates $(x_n^{a+},x_n^{a-})$,
$1\le n\le P$, $1\le a\le N$,  and taking 
\be \O=-\sum_{n=1}^P \sum_{a=1}^N
 \left( \frac{A_n^a(x_n^{a+})H^adx_n^{a+}}{(1+\l)^n} +
        \frac{B_n^a(x_n^{a-})H^adx_n^{a-}}{(1-\l)^n} \right), \ee
where $\{H^a\}$, $a=1,\ldots,N$ is a basis for the algebra of antihermitian,
diagonal $N\times N$ matrices. In this hierarchy, left dressings on $U_0$ 
correspond precisely to coordinate translations in the sector with
the scalar functions $A_n^a,B_n^a$ constant.

The physical or geometric significance of these PCM hierarchies remains
to be understood. 
An alternative approach to defining a PCM hierarchy was given in \c{b}.

\subsection{The Virasoro symmetry}\la{Virsym}

In this section we consider the symmetries of PCM associated with
reparametrisations of $U_0(\l)$. We consider the infinitesimal 
reparametrisations $U_0(\l)\rightarrow U_0(\l+\e_m\l^{m+1})$, where the
$\e_m$ are infinitesimal parameters and $m\in{\bf Z}$, or, equivalently,
variations $\delta U_0 = \e_m\l^{m+1} U_0'(\l)$. The
prime denotes differentiation with respect to $\l$.

These variations give rise to a centreless Virasoro algebra of
infinitesimal symmetries of PCM. In \c{s} Schwarz documents the 
existence of `half' of this algebra. Schwarz's symmetries
are associated with reparametrisations that fix the points $\l=\pm 1$.
We shall see that from a technical standpoint these are simpler to
handle than the full set of symmetries. But there is also a fundamental
reason to make such a restriction. If we were to consider finite 
reparametrisations, we would need to ensure that the contour $\C$
remains qualitatively unchanged.
The simplest way to do this is to require the points $\l=\pm 1$ to be 
fixed. In \c{u} Uhlenbeck identifies an $S^1$ symmetry of PCM. It is a 
simple exercise to check that this symmetry corresponds, in our formalism,
to global reparametrisations of the $\l$-plane fixing the points $\pm 1$,
i.e. M\"obius transformations of the form
\be \l \rightarrow \frac{a\l+b}{b\l+a},\quad\quad a^2+b^2=1.  \ee
At the level of infinitesimal symmetries, however, the need to fix $\pm 1$
is really superfluous, and so we find a full Virasoro algebra of
symmetries. But as we have said above, the symmetries fixing $\pm 1$ are 
technically easier, which is why Schwarz was able to identify
them, and also for the more general symmetries we can be quite certain
that there exists no exponentiation. 

With this introduction, we consider the variations
$\d_m U_0 = \e_m\l^{m+1} U_0'(\l)$. These manifestly realise the algebra
$[\d_m,\d_n]=(n-m)\d_{n+m}$. This realisation descends to the physical
fields. Using $U_0 = e^{-M}S^{-1}Y$ we have the chain of implications
\begin{eqnarray}
\delta_m U_0 &=&\e_m\l^{m+1} (-M'e^{-M}S^{-1}Y - e^{-M}S^{-1}S'S^{-1}Y
           +e^{-M}S^{-1}Y')\\
\delta_m U &=&e^M \delta_m U_0 \nonumber\\
         &=& \e_m\l^{m+1} (-M'S^{-1}Y - S^{-1}S'S^{-1}Y +S^{-1}Y')\\
\delta_m S &=&-(S\delta_m U Y^{-1})_-S \nonumber\\
         &=&-\e_m\left(\l^{m+1} (-SM'S^{-1} - S'S^{-1}+Y'Y^{-1})\right)_-S \\
\delta_m Y &=&(S\delta_m U Y^{-1})_+Y  \nonumber\\
         &=&\e_m\left(\l^{m+1} (-SM'S^{-1}+Y'Y^{-1})\right)_+Y.
\end{eqnarray}
In the last equation we have used the fact that for all $m$, 
$\l^{m+1}S'S^{-1}$ takes values in ${\cal G}_-$. Of the remaining two terms, 
the first has a  ${\cal G}_+$ piece
originating in the double pole of $M'$ at 
$\lambda=\pm 1$. To explicitly compute this is a simple exercise. 
For the second term, we use a contour integral formula for the 
projection. We thus arrive at the final result
\begin{eqnarray}
\delta_m Y~Y^{-1}&=&\e_m\left( \frac1{2\pi i}\int_{\C} 
  \frac{\mu^{m+1}Y'(\mu)Y^{-1}(\mu)}{\mu - \l}d\mu\right. \nonumber\\
              && 
+(-1)^m\left(s_0~{\textstyle{\int A}}~ s_0^{-1}\right) 
        \left( \frac1{(1+\l)^2}-\frac{m+1}{1+\l}\right)
+\frac{(-1)^m}{1+\l}\left[s_1s_0^{-1},\left(s_0~
   {\textstyle{\int A}}~ s_0^{-1}\right)\right]
    \nonumber\\
&& \left.
+\left(\wt s_0~{\textstyle{\int B}}~ \wt s_0^{-1}\right)
 \left( \frac1{(1-\l)^2}-\frac{m+1}{1-\l}\right)
+\frac1{1-\l}\left[\wt s_1\wt s_0^{-1},
   \left(\wt s_0~
   {\textstyle{\int B}}~ \wt s_0^{-1}\right)\right] \right)
\end{eqnarray}
Here $\int A$ and $\int B$ are shorthand for $\int_{x_0^+}^{x^+} A(y^+)dy^+$
and $\int_{x_0^-}^{x^-} B(y^-)dy^-$ respectively. The $g$ transformations are
read off by setting $\l$ to zero. 
In the expression for $\delta_mg$, the contour integral term is evaluated, 
depending on the value of $m$, by shrinking $\C$ to a contour around either
$0$ or $\infty$. Explicitly for the SL(2)
subalgebra of the Virasoro algebra, we obtain
(omitting the overall infinitesimal parameters),
\begin{eqnarray*}
g^{-1}\delta_{-1}g &=& \phi + (s_0{\textstyle{\int}} A s_0^{-1}) - 
      (\wt s_0{\textstyle{\int}} B \wt s_0^{-1})
    +\left[s_1s_0^{-1},(s_0{\textstyle{\int}} A s_0^{-1}) \right]
 -\left[\wt s_1 \wt s_0^{-1},(\wt s_0{\textstyle{\int}} B \wt s_0^{-1})\right]
                  \\
g^{-1}\delta_{0}g &=&  -
     \left[s_1s_0^{-1},(s_0{\textstyle{\int}} A s_0^{-1})\right]-
  \left[\wt s_1 \wt s_0^{-1},(\wt s_0{\textstyle{\int}} B \wt s_0^{-1})\right]
     \\
g^{-1}\delta_{1}g &=&  f- (s_0{\textstyle{\int}} A s_0^{-1})+
        (\wt s_0{\textstyle{\int}} B \wt s_0^{-1})+
    \left[s_1s_0^{-1}, (s_0{\textstyle{\int}} A s_0^{-1})\right]-
   \left[\wt s_1 \wt s_0^{-1},(\wt s_0{\textstyle{\int}} B \wt s_0^{-1})\right].
\end{eqnarray*}
We see that in these formulae,  not only do the leading coefficients
$s_0,s_1,\wt s_0,\wt s_1$ in the expansions of $S$
appear, but also the fields
$\phi$ and $f$, coefficients in the expansions of $Y$ around
$0$ and $\infty$ respectively (see section \ref{exts}).
The work required to check directly that these, or any of the $\delta_m$'s,
are symmetries is formidable, but we again emphasize that the advantage
of the present framework is that such direct checks are not necessary in order
to prove that the physical fields carry a representation of the full
centreless Virasoro algebra.

Schwarz \c{s} has previously found half a Virasoro algebra. 
We observe that if we define transformations
$\Delta_m=\delta_{m+1}-\delta_{m-1}$ a substantial simplification takes
place, yielding the formula
\be \Delta_m g = -\e_m g \left(
\frac1{2\pi i}\int_{\C} \mu^{m-1}(\mu^2-1)Y'(\mu)Y^{-1}(\mu)d\mu
  + 2(-1)^m(s_0~{\textstyle{\int}} A~s_0^{-1}) - 
      2(\wt s_0~{\textstyle{\int}} B~\wt s_0^{-1})
\right). \ee
We will see in section \ref{ABtr} (see equations \r{At},\r{Bt})
that the second and third terms in the
above expression are individually symmetries of PCM that 
mutually commute and commute with
all the symmetries being considered here. Removing these terms
gives exactly the `half-Virasoro' symmetries of \c{s}
\be \wt\Delta_m g = -\e_m g
\frac1{2\pi i}\int_{\C} \mu^{m-1}(\mu^2-1)Y'(\mu)Y^{-1}(\mu)d\mu,
\quad\quad m\in{\bf Z}. \ee
Thus we see the precise nature of Schwarz's symmetries as
combinations of reparametrisations preserving the points $\l=\pm 1$
with certain simple symmetries that act on the $A,B$ fields but leave
$U_0$ invariant. Taking the 
appropriate combinations we see that for the simplest Schwarz symmetry
$\wt \Delta_0$ 
\be g^{-1}\wt\Delta_0 g = \phi-f \ee
and using \r{Af} and \r{peq},
\be \wt\Delta_0 A_\pm = \mp 2 A_\pm+[A_\pm,f]. \ee
This is easily checked to be a symmetry. The 
symmetry $\Delta_0$ acts on the physical fields in a much more 
complicated way:
\begin{eqnarray*}
g^{-1}\Delta_0 g &=& \phi-f-2(s_0{\textstyle{\int}} A s_0^{-1})+
        2(\wt s_0{\textstyle{\int}} B \wt s_0^{-1}) \\
\Delta_0 A_+ &=& -4A_+ + [A_+,f] -2 \left[ [s_1s_0^{-1},A_+],
       (s_0{\textstyle{\int}} A s_0^{-1})  \right] + \left[A_+,
    (\wt s_0{\textstyle{\int}} B \wt s_0^{-1})\right]\\
\Delta_0 A_- &=& 4A_-+[A_-,f]+2\left[[\wt s_1\wt s_0^{-1},A_-], 
   (\wt s_0{\textstyle{\int}} B \wt s_0^{-1}) \right] - \left[A_-,
    (s_0{\textstyle{\int}} A s_0^{-1})  \right]. 
\end{eqnarray*}

\subsection{Transformations of the free fields $A(x^+),B(x^-)$}\la{ABtr}

Following the by now familiar reasoning,
an infinitesimal transformation $A(x^+)\mapsto A(x^+)+\delta A(x^+)$ 
induces the following transformations on $Y, g, A_+, $ $A_-$:
\begin{eqnarray*}
\delta Y &=& -\frac{(s_0~\int\delta A ~s_0^{-1})}{1+\l} ~ Y\\
\delta g &=&g~ (s_0~{\textstyle{\int}} \delta A~s_0^{-1}) \\
\delta A_+&=& s_0~\delta A~s_0^{-1} -
   \left[A_+, \left[s_1s_0^{-1}~,~(s_0 ~
   {\textstyle{\int}} \delta A~s_0^{-1})\right]\right] \\
\delta A_-&=& \left[ A_-~,~(s_0~{\textstyle{\int}}\delta A~s_0^{-1})\right].
\end{eqnarray*}
Here we have written $\int\delta A$ as shorthand for 
$\int_{x_0^+}^{x^+} \delta A(y^+) dy^+$. As expected, the spectrum of
$A_-$ remains invariant, while that of $A_+$ is shifted. Using the flow
equations for $s_0,s_1$, it is easy to check that these are genuine
symmetries, i.e. that $\p_-\delta A_++\p_+\delta A_-=0$.

There are a variety of possibilities for $\delta A(x^+)$. If $\{H^a\}$, 
$a=1\ldots N$, is a basis of the algebra of antihermitian diagonal matrices,
we can consider variations
$\delta A(x^+) \sim  (x^+ )^m H^a$, $a=1,\ldots,N$, $m\in{\bf Z}$. 
This gives a loop 
algebra of symmetries, corresponding to translations of $A(x^+)$. 
Taking $\delta A(x^+) \sim (x^+ )^m A'(x^+) $, $m\in{\bf Z}$, gives a 
centreless Virasoro algebra of symmetries, corresponding to reparametrizations
of $A(x^+)$. Taking $\delta A(x^+) \sim (x^+ )^m A(x^+)$, $m\in{\bf Z}$, 
gives an infinite 
dimensional abelian symmetry algebra corresponding to $x^+$-dependent
rescalings of $A(x^+)$. Clearly these symmetries are not independent: The
latter two families can be written in terms of the first family, but 
the generators are then {\em field dependent} combinations of the 
generators of the first family. 

Analogous sets of symmetries can be obtained from infinitesimal variations
of $B(x^-)$. 

The simple variation
$\delta A(x^+) = \epsilon A(x^+)$, where $\epsilon$ is a 
constant infinitesimal parameter, yields the symmetry
\be \delta g = \epsilon g (s_0~{\textstyle{\int}} A~s_0^{-1}),  
      \la{At}\ee
whereas the transformation $B\mapsto (1+\zeta)B$, where $\zeta$ is 
also an infinitesimal parameter, yields
\be \delta g = \zeta g (\wt s_0~
      {\textstyle{\int B}}~\wt s_0^{-1}).  \la{Bt}\ee
These  transformations were used in section \r{Virsym} to make contact 
between our Virasoro symmetries and those of \c{s}.

\section{Concluding remarks}

We have seen that formulating the nonlinear equations  of motion
\r{pcm} of the PCM in the form of the simple linear system \r{lin} makes
the precise nature of their integrability completely transparent. It
yields a novel free-field parametrisation of the space of solutions,
which we have used to classify all the symmetries of on-shell PCM
fields in terms of natural transformations on the free-field data.
The confusing cacophony of symmetry transformations in the literature
is thereby seen to arise in the most natural fashion imaginable. We
have thus demonstrated that this notion of complete integrability,
previously applied to traditional soliton systems, like the
KP, NLS and KdV hierarchies, encompasses the Lorentz--invariant
PCM field theories. We believe that this notion of integrability is
a universal one and we expect a clarification of the nature of the
integrability of the self-dual Yang-Mills and self-dual gravity 
equations by similarly reformulating the twistor constructions for these
systems. Indeed Crane \c{c} has already discussed a loop group of 
symmetries in terms of an action on free holomorphic data in twistor 
space. 

Our construction raises many questions. 

\noindent 1) Standard integrable soliton systems exhibit multiple
hamiltonian structures and infinite numbers of conservation laws,
both these phenomena being symptoms of their integrability. These 
phenomena ought to have a natural explanation in terms of the 
associated simple linear systems (free-field data). For the PCM,
some work on such structures exists \c{di}.

\noindent 2) The free-field parametrisation of solutions of PCM should
play a critical role in the quantisation of the theory. What is the
relation with standard quantisations? (The PCM can be quantised in
different ways, using either the field $f$ or the field  $g$ as 
fundamental,  giving different results \c{n}.) How are we to
understand quantum integrability?

\noindent 3) There is a large body of related mathematical work,
mostly focusing on the enumeration and construction of solutions of the
PCM in Euclidean space (for recent references see \c{g}). Most of
our formalism goes through for the case of Euclidean space, but
the reality conditions are different, and a little harder to handle. 
An important class of solutions are the  {\em unitons} \c{u,w}. These
correspond, up to the need for right dressings by $G_+$ elements, to
$Y$'s with finite order poles at one of the two points $\pm 1$, and 
regular elsewhere. We wonder: What are the corresponding $U_0$'s?
(The work of Crane on self-dual Yang-Mills \c{c} may have an analog.)
Is there a natural geometric understanding of our construction? Or
a relation with the constructions of \c{w} or \c{g}?

%\noindent 3) There is a large body of related mathematical work
%(see \c{g,u,w} and references therein). Is there a natural
%geometric interpretation of our construction?

\noindent 4) Is there a geometric interpretation of our PCM hierarchy?

\vskip.5cm

\noindent{\it Acknowledgments.}
We should like to thank Bernie Pinchuk and Larry Zalcman for discussions
on Runge's theorem.
One of us (CD) is happy to thank the Emmy Noether Mathematics Institute of
Bar--Ilan University for generous hospitality.

\section*{Appendix. The PCM hierarchy} \label{App}
In section \r{Ld} we have described a procedure to generate a PCM hierarchy.
In this appendix we illustrate this procedure by obtaining the 
simplest integrable extension of the PCM equation. We use the $\O$ given in
\r{nO} for $P=2$. Using $Z=(S\O S^{-1})_+$ we obtain the following form
for $Z$:
\[ Z=-\left(\frac{A_+dx_1^+}{1+\l} + 
 \left( \frac{B_+}{(1+\l)^2} + \frac{[C_+,B_+]}{1+\l}\right) dx_2^+
+ \frac{A_-dx_1^-}{1-\l} + 
 \left( \frac{B_-}{(1-\l)^2} + \frac{[C_-,B_-]}{1-\l}\right) dx_2^-
\right). \]
The six fields $A_+,\linebreak[0]B_+,\linebreak[0]C_+,\linebreak[0]A_-,
\linebreak[0]B_-,\linebreak[0]C_-$ are defined in terms of the coefficients
of $S$ and the free fields $A_1(x_1^+),A_2(x_2^+),B_1(x_1^-),B_2(x_2^-)$.
They depend on the four coordinates $x_1^+,x_2^+,x_1^-,x_2^-$ and
are constrained in virtue of their defining relations thus:
$A_+$ commutes with $B_+$, $A_-$ commutes with $B_-$ and the 
spectra of $A_+,B_+,A_-,B_-$ depend only on 
$x_1^+,x_2^+,x_1^-,x_2^-$ respectively. If 
we nevertheless ignore these constraints 
and simply substitute the above form for $Z$ into $dZ=Z\wedge Z$, we find:

\noindent 1. $[A_+,B_+]=[A_-,B_-]=0$.

\noindent 2. The following system of evolution equations for 
$A_+,B_+,A_-,B_-$:
\begin{eqnarray*}
\p_{2+}A_+ &=&  -{\textstyle\frac12}[A_+,[[B_+,C_+],C_+]]-
       [B_+,\p_{1+}C_+ + {\textstyle\frac12}[[A_+,C_+],C_+]] \\
\p_{1-}A_+ &=&  {\textstyle\frac12} [A_+,A_-] \\
\p_{2-}A_+ &=&  {\textstyle\frac12} [A_+,{\textstyle\frac12}B_-+[C_-,B_-]]\\
\p_{1+}B_+ &=&  [B_+,[A_+,C_+]] \\
\p_{1-}B_+ &=&  {\textstyle\frac12} [B_+,A_-] \\
\p_{2-}B_+ &=&  {\textstyle\frac12} [B_+,{\textstyle\frac12}B_-+[C_-,B_-]] \\
\p_{1+}A_- &=&  {\textstyle\frac12} [A_-,A_+]\\
\p_{2+}A_- &=&  {\textstyle\frac12} [A_-,{\textstyle\frac12}B_++[C_+,B_+]]\\
\p_{2-}A_- &=&   -{\textstyle\frac12}[A_-,[[B_-,C_-],C_-]] -
        [B_-,\p_{1-}C_-+{\textstyle\frac12}[[A_-,C_-],C_-]]\\
\p_{1+}B_- &=&  {\textstyle\frac12} [B_-,A_+]\\
\p_{2+}B_- &=&   {\textstyle\frac12} [B_-,{\textstyle\frac12}B_++[C_+,B_+]]\\
\p_{1-}B_- &=&   [B_-,[A_-,C_-]] 
\end{eqnarray*}
These evidently 
imply that the spectra of $A_+,B_+,A_-,B_-$ depend only
on  $x^{1+},x^{2+},x^{1-},x^{2-}$  respectively, as required.

\noindent 3. The following evolution equations for $C_+,C_-$:
\begin{eqnarray*}
\p_{1+}C_- &=& -{\textstyle\frac14} A_+ - {\textstyle\frac12}[A_+,C_-]\\
\p_{2+}C_- &=& -{\textstyle\frac18}B_+ 
               +{\textstyle\frac14}([C_-,B_+]-[C_+,B_+]) 
               +{\textstyle\frac12}[C_-,[C_+,B_+]] \\
\p_{1-}C_+ &=& -{\textstyle\frac14} A_- - {\textstyle\frac12}[A_-,C_+]\\
\p_{2-}C_+ &=& -{\textstyle\frac18}B_- 
               +{\textstyle\frac14}([C_+,B_-]-[C_-,B_-]) 
               +{\textstyle\frac12}[C_+,[C_-,B_-]] .
\end{eqnarray*}
(In fact, from the $dZ=Z\wedge Z$ equation, both of the $C_-$ evolutions 
appear commutated with $B_+$ and both of the $C_+$ evolutions appear
commutated with $B_-$.)

This system is a 4-dimensional integrable system, but
its physical or geometric interpretation is not immediately apparent.
It has a variety of interesting reductions apart from
the reduction to PCM by setting $B_-=B_+=0$. We can consistently reduce
by taking $A_-=B_-$ or $A_+=B_+$ or both. Or we can take just $B_-=0$
(or $B_+=0$) in which case the $x_2^-$ (or $x_2^+$) dependence becomes 
trivial. For all these reductions, and the full system as well, the methods
of this paper give a free-field parametrisation of solutions.

%%%%%%%%%%%%%%%%%%%%%%%%%%%%%%%%%%


\begin{thebibliography}{99}
\bibitem{b} Bruschi, M., Levi, D., Ragnisco, O.: The chiral field
hierarchy. Phys.Lett. {\bf 88A}, 379-382 (1982).
\bibitem{g} Burstall, F.E.,
     Guest, M.A.: Harmonic two-spheres in compact symmetric
    spaces, revisited. Preprint (1996).
\bibitem{c} Crane, L.: 
Action of the loop group on the self-dual Yang-Mills equation.
Commun.Math.Phys. {\bf 110}, 391 (1987).
\bibitem{Chand} Devchand, C., Fairlie, D.B.: 
A generating function for hidden symmetries of chiral models. 
Nucl.Phys. {\bf B194}, 232-236 (1982).
\bibitem{di} Dickey, L.A.: Symplectic structure, Lagrangian, and
involutiveness of first integrals of the principal chiral field equation.
Commun.Math.Phys. {\bf 87}, 505-513 (1983).
\bibitem{d} Dolan, L.: Kac-Moody algebra is hidden symmetry of
chiral models. Phys.Rev.Lett. {\bf 47}, 1371-1374 (1981).
\bibitem{f} Feller, W.: An introduction to probability theory and its
applications, Volume I. New York, Chichester, Brisbane, Toronto: 
  John Wiley \& Sons, 1967.
\bibitem{haa} Haak, G., Schmidt, M., Schrader, R.: Group Theoretic
 Formulation of the Segal-Wilson Approach to Integrable Systems with 
 Applications.  Rev.Math.Phys. {\bf 4}, 451-499 (1992).
\bibitem{h} Harnad, J., Saint-Aubin, Y., Shnider, S.: Superposition 
of solutions to B\"acklund transformations for the SU(N) principal
$\sigma$-model. J.Math.Phys. {\bf 25}, 368-375 (1983). Quadratic 
psuedopotentials for Gl($N,\CC$) principal sigma models. Physica
{\bf 10D}, 394-412 (1984). B\"acklund transformations for nonlinear
sigma models with values in Riemannian symmetric spaces. 
Commun.Math.Phys. {\bf 92}, 329-367 (1984). The soliton correlation
matrix and the reduction problem for integrable systems. 
Commun.Math.Phys. {\bf 93}, 33-56 (1984).
\bibitem{l} Leznov, A.N.: B\"acklund transformation for main chiral
   field problem with an arbitrary semisimple algebra. Preprint (1991).
\bibitem{m} Mulase, M.: Complete integrability of the
   Kadomtsev-Petviashvili equation. Adv.Math. {\bf 54}, 57-66 (1984).
  Solvability of the super KP equation and a generalization 
  of the Birkhoff decomposition. Inv.Math. {\bf 92}, 1-46 (1988).
\bibitem{n} Nappi, C.R.: Some properties of an analog of the 
nonlinear sigma model. Phys.Rev. {\bf D21}, 418-420 (1980).
\bibitem{o} Ogielski, A.T., Prasad, M.K., Sinha, A. Chau Wang, L-L.:
B\"acklund transformations and local conservation laws
for principal chiral fields. Phys.Lett. {\bf 91B}, 387-391 (1980).
\bibitem{p} Pohlmeyer, K.: Integrable Hamiltonian systems and
interactions through quadratic constraints. Commun.Math.Phys. {\bf 46},
207-221 (1976).
\bibitem{r} Rudin, W.: Real and complex analysis. New York, St.Louis,
  San Francisco, Toronto, London, Sydney: McGraw-Hill Book Company (1966).
\bibitem{j} Schiff, J.: Symmetries of KdV and loop groups. Preprint (1996).
   (Archive number solv-int/9606004.)
\bibitem{s} Schwarz, J.H.: Classical symmetries of some two-dimensional
models. Nucl.Phys. {\bf B447}, 137-182 (1995).
\bibitem{un} Ueno, K., Nakamura, Y.: The hidden symmetry
of chiral fields and the Riemann-Hilbert problem.
Phys.Lett. {\bf 117B}, 208-212 (1982).
\bibitem{u} Uhlenbeck, K.: Harmonic maps into Lie groups (classical 
solutions of the chiral model). J.Diff.Geom. {\bf 30}, 1-50 (1989).
\bibitem{w} Ward, R.S.: Classical solutions of the chiral model, unitons,
and holomorphic vector bundles. Commun.Math.Phys. {\bf 128}, 319-332 (1990).
\bibitem{zm} Zakharov, V.E., Mikhailov, A.V.: Relativistically invariant
two-dimensional models of field theory which are integrable by means
of the inverse scattering problem method. Zh.Exp.Teor.Fiz. {\bf 74} 1953-1973
(1978) (English translation: Sov.Phys.JETP {\bf 47}, 1017-1027 (1978)).

\end{thebibliography}
\end{document}